\documentclass[manuscript,screen]{acmart}

\AtBeginDocument{%
  \providecommand\BibTeX{{%
    \normalfont B\kern-0.5em{\scshape i\kern-0.25em b}\kern-0.8em\TeX}}}

\setcopyright{none}

\usepackage{diagbox}
\usepackage{multirow}
\usepackage{hyperref}
\hypersetup{
    colorlinks=true,
    linkcolor=blue,
    filecolor=magenta,      
    urlcolor=cyan,
    pdftitle={Overleaf Example},
    pdfpagemode=FullScreen,
    }

\begin{document}

\title[Exploring Musical Roots: Influence Attribution for a Generative Music Model]{Exploring Musical Roots: Applying Audio Embeddings to Empower Influence Attribution for a Generative Music Model}

\author{Julia Barnett}
\orcid{0000-0002-3476-1110}
\affiliation{%
  \institution{Northwestern University}
  \city{Evanston}
  \state{IL}
  \country{USA}
}
\email{JuliaBarnett@u.northwestern.edu}

\author{Hugo Flores Garcia}
\orcid{0009-0004-7302-4203}
\affiliation{%
  \institution{Northwestern University}
  \city{Evanston}
  \state{IL}
  \country{USA}
}
\email{hugofg@u.northwestern.edu}

\author{Bryan Pardo}
\orcid{0000-0002-1427-6492}
\affiliation{%
  \institution{Northwestern University}
  \city{Evanston}
  \state{IL}
  \country{USA}
}
\email{pardo@northwestern.edu}

\renewcommand{\shortauthors}{Barnett et al.}
\begin{abstract}
  Every artist has a creative process that draws inspiration from previous artists and their works. Today, ``inspiration'' has been automated by generative music models. The black box nature of these models obscures the identity of the works that influence their creative output. As a result, users may inadvertently appropriate, misuse, or copy existing artists' works. We establish a replicable methodology to systematically identify similar pieces of music audio in a manner that is useful for understanding training data attribution. A key aspect of our approach is to harness an effective music audio similarity measure. We compare the effect of applying CLMR \cite{spijkervet2021contrastive} and CLAP \cite{clap_wu_2023} embeddings to similarity measurement in a set of 5 million audio clips used to train VampNet \cite{garcia2023vampnet}, a recent open source generative music model. We validate this approach with a human listening study. We also explore the effect that modifications of an audio example (e.g., pitch shifting, time stretching, background noise) have on similarity measurements. This work is foundational to incorporating automated influence attribution into generative modeling, which promises to let model creators and users move from ignorant appropriation to informed creation. Audio samples that accompany this paper are available at \url{https://tinyurl.com/exploring-musical-roots}.

\end{abstract}




\received{22 January 2024}
\maketitle

\section{Introduction}

Generative models have revolutionized the ease with which anyone can create polished works of art. 
With today's models there is an opaque nature to the generation process---it is never clear to the end user what data influences and shapes their newly crafted essay from ChatGPT \cite{openai_2022}, digitized surrealist art from DALLE-2 \cite{ramesh2022hierarchical}, or soulful jazz in the style of Rihanna from MusicLM \cite{agostinelli2023musiclm}. Even further, due to the vast amounts of  data they were trained on, it is usually not even clear when these models are ``creating'' near replicas of existing items from their training data.

For users of generative models to be informed and responsible creators, 
there needs to be a mechanism that provides information about works in the model's training data that were highly influential upon the generated output, or directly copied by the model. This would allow the user to both cite existing work and learn about the influences of their generated output. We assume a model-generated product that is a copy or near-copy of a work in the model's training set indicates the model was influenced by that work. To develop methods to automatically detect the influences upon model-generated products it is, therefore, essential to develop good measures of similarity between works.    

In text, it is straightforward to detect when language models copy strings of text verbatim, given access to the training data. There is a growing body of work quantifying the degree to which these large language models memorize training data \cite{feldman2020neural, carlini2021extracting, carlini2022quantifying}. 
In the image space, it is more complex due to the high-resolution multi-pixel outputs of models, but work is being done to detect ``approximate memorization'' by finding highly similar images from the training data \cite{somepalli2023diffusion, carlini2023extracting}. In this work, we perform a similar analysis in the space of music audio generation, developing a methodology and framework for finding approximate matches that has been validated in a human listening study and for which the effect of a variety of audio perturbations has been systematically studied.

Extending from the image space, we define a measure of approximate memorization in deep generative audio models by establishing a threshold for high similarity and memorization of training data against a large repertoire of 5,000,000+ song clips. We take inspiration from the ``split-product’’ measure for image similarity  from Somepalli et al. \cite{somepalli2023diffusion}, which breaks the embedded feature vector of images into smaller chunks to compare inner products of corresponding localized features. In our work, every audio file is split into 3-second segments (a.k.a. clips), each of which is encoded as a feature vector (either a CLMR \cite{spijkervet2021contrastive} or CLAP \cite{clap_wu_2023} embedding) produced by a machine learning model trained to encode audio for the purpose of measuring similarity. See Section \ref{sec:embeddings} for details. We measure similarity between generated clips and training data clips to find similarity between sub-portions of songs (e.g., a single musical phrase), returning the songs with the most similar clips. We also 
 evaluate the extent to which similarity measured in this way agrees with similarity assessments by human listeners (Section \ref{sec:experiment}). 

We apply our approach to a music audio generation model that is representative of a widely-used class of generative models. Perhaps the most commonly-applied method to music audio generation is language-model-style generation, where audio is converted into tokens and then a transformer architecture is trained to generate token streams using masked prediction. AudioLM \cite{borsos2023audiolm}, Jukebox \cite{dhariwal2020jukebox}, MusicGen \cite{copet2023simple}, MusicLM \cite{agostinelli2023musiclm}, StemGen \cite{parker2023stemgen}, SoundStorm \cite{borsos2023soundstorm}, and VampNet \cite{garcia2023vampnet} all use this approach. We validate our framework on VampNet, a model trained on 795k music songs collected from the internet. Since it is open source, the model's weights are publicly available, and the authors were able to provide specific information on the content of the training data. Aside from Meta's MusicGen \cite{copet2023simple}, for which we do not have access to the training data, it is the only model for generating music audio that is publicly available at the time of this writing. While we apply our work to VampNet, our evaluation framework is both model and training data agnostic, and we expect it will be useful for both model creators to assess the degree to which their models memorize and for end users to police themselves as they continue to generate ``new’’ pieces of music.

This paper makes the following \textbf{key contributions}. Primarily, it establishes \textbf{an easily replicable methodology and framework to perform training data attribution for a generative music model} (Section \ref{sec:data_and_method_big_section}), which has been validated in a human-listener study (Section \ref{sec:experiment}). Second, \textbf{we systematically explore the robustness of embedding-based similarity measures for music audio (CLMR and CLAP) to audio perturbations such as pitch shift, time stretch, and mixture with different types of noise} (Section \ref{sec:perturbation}). We do this because generative models, even when creating near-copies of training data, are likely to add some form of variation to the outputted generation, making it essential to understand how robust this methodology is to such anticipated perturbations.

Our formal research questions are:
\begin{enumerate}
    \item How can we systematically identify similar pieces of music from training data to new generations in a manner that is useful for understanding training data attribution of newly generated music audio?
    \item How do different perturbation types and amounts affect the ability of the evaluated similarity measure(s) to quantitatively identify similar pieces of music? 
\end{enumerate}

This work provides the basis for creators and users of generative models to have informed attribution of influence. Exposing this information gives creators the chance to appropriately cite influence and, perhaps even more importantly, learn about the influences upon the created work, transforming the generative model from a crutch that replaces artistic knowledge to a scaffold towards becoming a better artist. Moving forward, automated attribution of influence is a key component necessary to appropriate compensation for creators of training data, based on their degree of contribution to a generated work.

\section{Relevant Literature}
\subsection{Memorization in Generative Language Models}
It is well established that large language models (LLMs) applied to text are capable of memorizing part of their training data \cite{carlini2019secret, feldman2020neural, carlini2021extracting, hu2022membership, peng2023near, de2023evaluation, tirumala2022memorization}. In fact, it was recently established that large language models like the 6 billion parameter GPT-J model memorize at least 1\% of the training data \cite{carlini2022quantifying}. Though there are potential benefits to memorization such as preventing the ``hallucination’’ of plausible sounding facts \cite{biderman2023emergent}, this raises a number of concerns, foremost amongst which are sensitive data leaks and copyright infringement. In response, there have been efforts to predict exactly which sequences of text that these models will memorize \cite{biderman2023emergent}. If access to the training data is available, it is relatively straightforward to detect when language models copy strings of text verbatim due to the ability to check for exact sequences of tokens. It becomes more complex in the image and audio space due to the complex nature of the data.  

\subsection{Memorization in Generative Image Models}

Images created by generative models pose the same set of risks as memorized training data from text LLMs---sensitive data leaks, copyright infringement---possibly to an even larger degree due to the difficulty of catching issues. With respect to images, memorization and duplication of neural models are fundamentally different than a text-based language model; instead of memorizing and reproducing items verbatim from the training data, they create images sufficiently similar to warrant content replication \cite{somepalli2023diffusion}. Using the same metric for memorization of text and images is analogous to asserting an identical string of text is not memorized if it is outputted in italics instead of standard typeset. 

Carlini et al. \cite{carlini2023extracting} propose an approximation of a distance metric for memorization in this complex image space. A generated image whose nearest neighbor in the training data falls closer than a determined threshold ($\delta$), when embedded in the appropriate manifold, is labeled as a memorized example even if it is not a verbatim copy. Somepalli et al. \cite{somepalli2023diffusion} demonstrate that diffusion models replicate images from training data with high fidelity, setting a lower bound for memorization of Stable Diffusion at 1.88\% of the time \cite{rombach2022high}. We extend the methodology from Somepalli et al. \cite{somepalli2023diffusion} and Carlini et al. \cite{carlini2023extracting} into the audio domain for the purpose of this paper. 


\subsection{Audio Retrieval and Music Similarity}

\subsubsection{Prior Work in Music Similarity and Memorization in Generative Audio Models}

Audio similarity measurement is similarly difficult to image similarity measurement. Given a recorded melody in the training data, we would wish to identify a generated music file as similar if it is the same melody, despite being played at a different speed, or in a higher key, or with background noise. 

Detecting audio similarity is by no means a new concept. Popularized in the early 2000s, audio fingerprinting \cite{cano2005review, haitsma2002highly} aims to detect exact copies of a given piece of audio. In 2006, Shazam popularized this method for the general public with a system utilizing query-by-example to put audio fingerprinting in the hands of everyday users \cite{wang2006shazam}. Traditional audio fingerprinting (e.g., Shazam \cite{wang2003industrial}) depends on low-level structural details that are not typically regenerated by generative models, so it is not a relevant approach for this methodology. 

Prior and surrounding this was ``query by humming’’ \cite{kotsifakos2012survey} or ``query by vocal imitation'' where people would try to identify an audio recording or song by humming it to the best of their ability. More recently, vocal query by example has been handled by embeddings that place vocal imitations and original audio into a space where they are directly comparable \cite{zhang2020vroom}. Selecting good embeddings to identify similarity is a key component of this work. Coversong ID has also been a popular method of detecting audio similarity---covers can differ in timbre, tempo, key, and more, making it a difficult music retrieval task studied in the early 2000s \cite{serra2010audio} and revisited with transformer architectures \cite{liu2023coverhunter}. 

Most of the (limited) work examining similarity of audio made by generative models has been in the context of a different purpose, rather than the focus of an in-depth exploration. Examples include creating new strategies for text-to-music generation in order to create more novel songs \cite{chen2023musicldm} or brief ad-hoc memorization evaluations of a model at time of release \cite{agostinelli2023musiclm, copet2023simple}. Perhaps the closest work to our own is by Bralios et al. \cite{bralios2023generation}. They recently examined replication of audio utilizing text-to-audio latent diffusion models. Their work used the AudioCaps \cite{kim2019audiocaps} set of 46K examples of general audio sounds, such as explosions or people cheering. They define replication of training data as ``containing nearly-identical complex spectro-temporal patterns.'' They did not perform any subjective evaluation by human listeners to validate their approach to measuring similarity. Our work instead focuses on the generative music domain, uses a much larger dataset of 5 million song clips, and has a primary goal of being easily replicable by any model creator. We also explore the use of multiple audio embeddings, validate our similarity measures with a human listening study, and systematically explore the effect of audio perturbations on similarity scores, none of which were done in the prior work. 


\subsubsection{Measuring Audio Similarity with Embeddings}

The key to measuring similarity effectively is to have a representation that highlights the task-relevant features. Most popular right now in the age of generative modeling---and the method employed by this work---is measuring audio similarity with embeddings. Audio embeddings are continuous vector representations for excerpts of audio that that can be used as a proxy measure of audio similarity between two music excerpts. These embeddings are usually retrieved from the internal representations of a deep neural network model trained on a proxy task like generative pre-training \cite{castellon2021codified}, contrastive learning \cite{clap_wu_2023, spijkervet2021contrastive}, audio classification \cite{kim2019improving}, autoencoding \cite{defossez2022high, kumar2023high}, among other methods \cite{li2022map, cramer2019look}. 

To use an audio embedding model to measure the similarity of a collection of audio excerpts, we pass the audio signals through the embedding network, which gives us a multi-dimensional vector output for each audio signal. This vector output is commonly referred to as an ``audio embedding.'' To obtain a list of the most similar audio signals for a given query audio signal, we extract the embeddings for each audio signal using an audio embedding model of our choice. We then compute a cosine or L1 distance between our query audio signal and the audio signals in the database, returning a ranked list, where audio signals with a higher similarity to the query audio are ranked higher in the list. 

The choice of audio embedding model can have a large impact on the results. There are a variety of embeddings capturing different features of audio: contrastive language-audio pretraining (CLAP) \cite{clap_wu_2023}, contrastive learning of musical representations (CLMR) \cite{spijkervet2021contrastive}, codified audio language modeling (CALM) \cite{castellon2021codified}, encodec \cite{defossez2022high}, music2vec \cite{li2022map}, Open L3 \cite{cramer2019look}, and VGGish \cite{kim2019improving} to name a few. These methods do not inherently scale (e.g., to 100 million songs in the Apple music library) and may require a method like locality-sensitive hashing \cite{jafari2021survey} to improve efficiency of searching. 

After experimenting with multiple options, we focus on CLAP \cite{clap_wu_2023} and CLMR \cite{spijkervet2021contrastive} embeddings for this work. These both are state-of-the-art technology as well as produce meaningful similarity in our own analysis in terms of human validated similarity through listening tests, robustness to perturbation, and ability to return relevant top songs.

\section{Data and Methodology}\label{sec:data_and_method_big_section}

\subsection{Scope of Analysis}\label{sec:scope}

We want to create a system that identifies music that is both quantitatively and subjectively similar to human ears. We do not focus on any individual specific feature of music for this similarity metric (e.g., timber, melody, style, rhythm), but rather use one of two possible embedding approaches (CLAP or CLMR) to encode audio and then examine whether similarity in these embedding spaces aligns with human subjective evaluation (Section \ref{sec:experiment}). We invite the reader to listen audio examples at \url{https://tinyurl.com/exploring-musical-roots}.

One obvious feature of music similarity and memorization that is copyrightable is lyrics. We are purposefully excluding lyrics from this analysis because they fall more in line with the text domain, which has been extensively studied. We instead focus our scope on the overall sound of these musical generations.

This framework is not meant for the legal domain---we are not proposing a tool to systematically identify copyright infringements. We acknowledge that there are three elements of music that can be copyrightable: (1) lyrics, (2) individual mastered recordings, and (3) unique elements of the composition \cite{lund2013fixing}. Copyright law in regard to generative models is a hot topic  \cite{arewa2005jc, samuelson2023generative, reichman1997intellectual}, and we are not seeking to be the answer. We are instead focused on enabling creators of these models to systematically evaluate their own models to understand training data attribution.

\subsection{Data and Models Used}

Though our approach is model agnostic, we validate our framework on VampNet \cite{garcia2023vampnet}, a generative model trained on 795k songs collected from the internet. VampNet takes a masked acoustic token modeling approach to music audio generation. In the first stage, a Descript Audio Codec (DAC) \cite{kumar2023high} learns to encode the audio data in a discrete vocabulary of ``tokens''. VampNet then is trained to model the sequence of tokens. To create a playable audio file, the token sequence is converted back into the input domain via the DAC decoder. VampNet adopts a masked generative modeling approach with a parallel iterative decoding procedure, inspired by generative models that work in the visual domain, such as MaskGIT \cite{chang2022maskgit}. 
To control generation output with VampNet, conditioning is through example audio, either with a prefix (prompting the model to generate a continuation), postfix (prompting it to make an introduction) or as infill (masking a region in the middle of an audio file).

We choose VampNet to establish the feasibility of our method due to it being open source, the model's weights being publicly available, and the authors' willingness to provide specific information on the content of the training data. From a stylistic perspective, VampNet is especially suited to our task of attributing various pieces of training data to generated pieces of audio due to the positioning of VampNet as a ``powerful music co-creation tool'' being capable of maintaining style, genre, and high-level aspects of the music. We denote musical outputs of VampNet as ``vamps.'' 

While we apply our work to VampNet, our evaluation framework is both model and training data agnostic, and we expect it will be useful for both model creators to assess the degree to which their models memorize and for end users to police themselves as they continue to generate ``new'' music.

\subsection{Methodology}\label{sec:Methodology}

The primary goal of this work is to create an easily-replicable methodology for creators and users of a generative music model to approach training data attribution. We describe this methodology in this section.

\subsubsection{Similarity Metric}

Extending from the image space, we define a measure of approximate memorization in deep generative audio models by establishing a threshold for high similarity and memorization of training data against a large collection of 5 million 3-second song clips drawn from the songs in the training data of the VampNet generative model  \cite{garcia2023vampnet} for music audio. We take inspiration from the ``split-product'' measure for image similarity from Somepalli et al. \cite{somepalli2023diffusion}, which breaks the embedded feature vector of images into smaller chunks to compare inner products of corresponding localized features. In our work, every audio file is split into 3-second clips, each of which is encoded as a feature vector (i.e., an embedding produced by an appropriately selected model). We measure the cosine similarity between generated segments and training data segments to find similarity between sub-portions of songs (e.g., a single musical phrase), returning the most similar songs and the corresponding metadata.  

\subsubsection{Embeddings} \label{sec:embeddings}
At the core of this work is our ability to identify meaningful perceptually-similar songs. This is done by calculating the cosine similarity between two clips encoded as audio embeddings. 
The way those embeddings are created is critical to how we can evaluate them. 

After experimenting with multiple options we focus on two main embeddings: (1) contrastive learning of musical representations (CLMR) embeddings \cite{spijkervet2021contrastive}, which are particularly effective generic representations of musical audio that generalize well to new music datasets, and (2) contrastive language-audio pretraining (CLAP) embeddings \cite{clap_wu_2023}, which are state-of-the-art in both text-to-audio retrieval and music classification tasks. We use both CLAP and CLMR embeddings because they can be applied to any dataset of raw music audio without the need for any transformation or fine-tuning, they generalize well to out-of-domain datasets, and can be used as a baseline across different models and different genres. One goal for this framework is to be replicable and simple to implement. Therefore, utilizing these publicly available embeddings that generalize to any dataset is helpful to encourage adoption.


We upload all of the embeddings and their corresponding musical metadata to a vector database that allows us to quickly and efficiently search through millions of embeddings and return the top $k$ similar songs by a chosen similarity metric (e.g., cosine similarity) in milliseconds. More details about the engineering behind this architecture can be found in the Appendix \ref{sec:appendix-vector-database}.

\section{Listening Test: Experimental Design} \label{sec:experiment}

We are interested in providing users of generative models information about the potential influences on the output of the model. Our approach is to automatically measure similarity between the model's output and its training data at the time of generation. Presumably, output that is highly similar to a training audio clip was influenced by that clip. Of course, similarity is in the ear of the listener and many similarity measures may be created that do not align with human opinions. A primary goal of this work is building a replicable framework that will not require other audio researchers to conduct costly and cumbersome human listening tests. Therefore, we conduct an experiment with human listeners to demonstrate the alignment of our quantitative technique with human listening. We do so by utilizing ReSEval, a helpful framework that enables us to build subjective evaluation of audio tasks we can deploy on crowdworker platforms \cite{morrison2022reproducible}.

\subsection{Dataset Preparation} \label{subsec:dataset_preparation}
We encode all clips in the dataset using both CLAP \cite{clap_wu_2023} and CLMR \cite{spijkervet2021contrastive} embeddings. The objective similarity measure we use is cosine similarity of audio embeddings. For our methodology we use a ``split-product’’ approach (see Section \ref{sec:Methodology}). Each song is represented by a set of 3-second clips and similarity measurement is performed at the clip level. Thus in our experiment, we seek to validate the similarity at a 3-second level and present listeners  with song excerpts of 3 seconds in length.

To create the data for our study we take a random sample of 1,000 3-second clips from a dataset of 5 million clips drawn from songs used to train VampNet. For each of these 1,000 clips, we rank its top 10,000 closest clips in the dataset by cosine similarity. This is done independently for both CLAP and CLMR embeddings. To give a feel for the resulting range of values, the average closest \#5 song had a similarity score of 0.912 for CLAP; 0.872 for CLMR, and the average closest \#100 song had a score of 0.878 for CLAP; 0.800 for CLMR.

For each embedding (CLAP and CLMR), we fit a Gaussian to the distribution of similarity scores of the top 10,000 clips (see histograms in Table \ref{tab:human_evaluation_results}). We then segment the data into 4 bins. To create meaningful bins, we take the median cosine similarity of the top 10,000 (CLAP: 0.815; CLMR: 0.693), +1 standard deviation (CLAP: 0.885; CLMR: 0.784), and +2 standard deviations (CLAP: 0.955; CLMR: 0.875). Presumably, the more standard deviations above the median a similarity score is, the more similar the clips are. We use these bin centers to create bins $\pm 0.02$ for each of these similarity scores. To have a bin that represents a truly random song pair, we then take two random subsets of 1,000 clips from the full 5 million clip dataset and measure pairwise cosine similarity. The median similarity of this distribution (CLAP: 0.513; CLMR: 0.151) gives the similarity score one can expect between truly random song pairings. 

These bins can be seen in Table \ref{tab:human_evaluation_results}. We concentrate on the higher cosine-similarity values (drawn from the top 10k matches), since in this work we only seek to evaluate clips that have high similarity scores---we care much less about the differences between clips that are of lower similarity. Additionally, prior work has established that both CLAP and CLMR embeddings encode gross differences well; therefore, it is unlikely that low-similarity examples (e.g., $\le0.513$ for CLAP and $\le0.151$ for CLMR) will sound similar to human listeners.

\subsection{ABX Trials}
The gold-standard for perceptual audio studies is a lab-based listener study. These, however, are difficult and time-consuming. 
Cartwright et al. \cite{cartwright2018crowdsourced, cartwright2016fast} overcame this burden utilizing pairwise comparison performed over the web in a micro-task labor market (e.g., Amazon Mechanical Turk), duplicating the findings of a lab-based test. We leverage these findings and employ a pairwise comparison study design, performing the study on Mechanical Turk.

In our study, listeners are asked to perform ABX trials. In each trial, a target audio clip (X) is presented, along with two other clips (A and B). The listener is asked to rate which clip (A or B) is more like the target X. The proportion of listeners that find A more similar than B is an estimate of the probability that people find A more similar to X than B. 
 
Given a clip X that was drawn from that random sample of 1,000 clips, one can then create a pair of examples A and B by selecting them randomly from different bins (see Section \ref{subsec:dataset_preparation}). This lets us create ABX trials with known differences in cosine similarity to X between the paired examples A and B. We can then collect statistics on the probability that users will call A more similar to X than B is to X. The greater the difference in cosine similarity, the more skewed we expect the listening results to be. If true, our objective measure's similarity rankings align with human rankings.

We have 4 bins, resulting in 6 different pair-wise comparisons (bin 1 vs. bin 2, 1 vs. 3, 1 vs. 4, 2 vs. 3, 2 vs. 4, and 3 vs. 4). To have 150 evaluations per bin (900 evaluations total), we need 90 people to listen to 10 ABX comparisons each. We randomly choose 15 prompt ``X'' clips from the training data, with their respective 4 clips within the bins chosen as detailed above for the A and B comparison. On average, each song has 10 evaluations per bin. 

Appendix \ref{sec:reseval-example} contains an example of what an evaluator would see while performing the ABX task. An example set of clips for an ABX evaluation is at \url{https://tinyurl.com/exploring-musical-roots}.

\subsection{Participant Recruitment}

For this experiment we utilized Amazon Mechanical Turk (MTurk). We had 150 participants evaluate similarity scores of CLAP embeddings and 150 participants evaluate similarity of CLMR embeddings. After ensuring participants passed the listening tests, we collected at least 60 evaluations of each set of ABX trials, and on average had 10 evaluations of each of the 6 bin comparisons per clip. We paid each evaluator to annotate 1 set of 10 ABX trials, which takes approximately 4 minutes. We paid \$1.50 for each completed set (an estimated pay of \$22.50 per hour). We recruited US residents on MTurk with an approval rating of at least 98 and 1,000 approved tasks. We filtered out bots 
by excluding evaluations that failed a pre-screening listening test. Evaluators failing the listening test were blocked from further evaluations. 


\subsection{Results}

\begin{table}[t]
\begin{tabular}{@{}cc|c|c|c|c@{}}
   \cmidrule(lr){1-6}
 \multicolumn{6}{c}{\textbf{Human Evaluation Results: ABX Listening Test}} \\
 \cmidrule(lr){1-6}
 \multirow{2}{*}{CLAP: Histogram of Top 10k Similar Clips}{} & \multicolumn{5}{c}{\textbf{CLAP Embeddings}} \\
   \cmidrule(lr){2-6}
\multirow{8}{*}{\includegraphics[height=3.3cm]{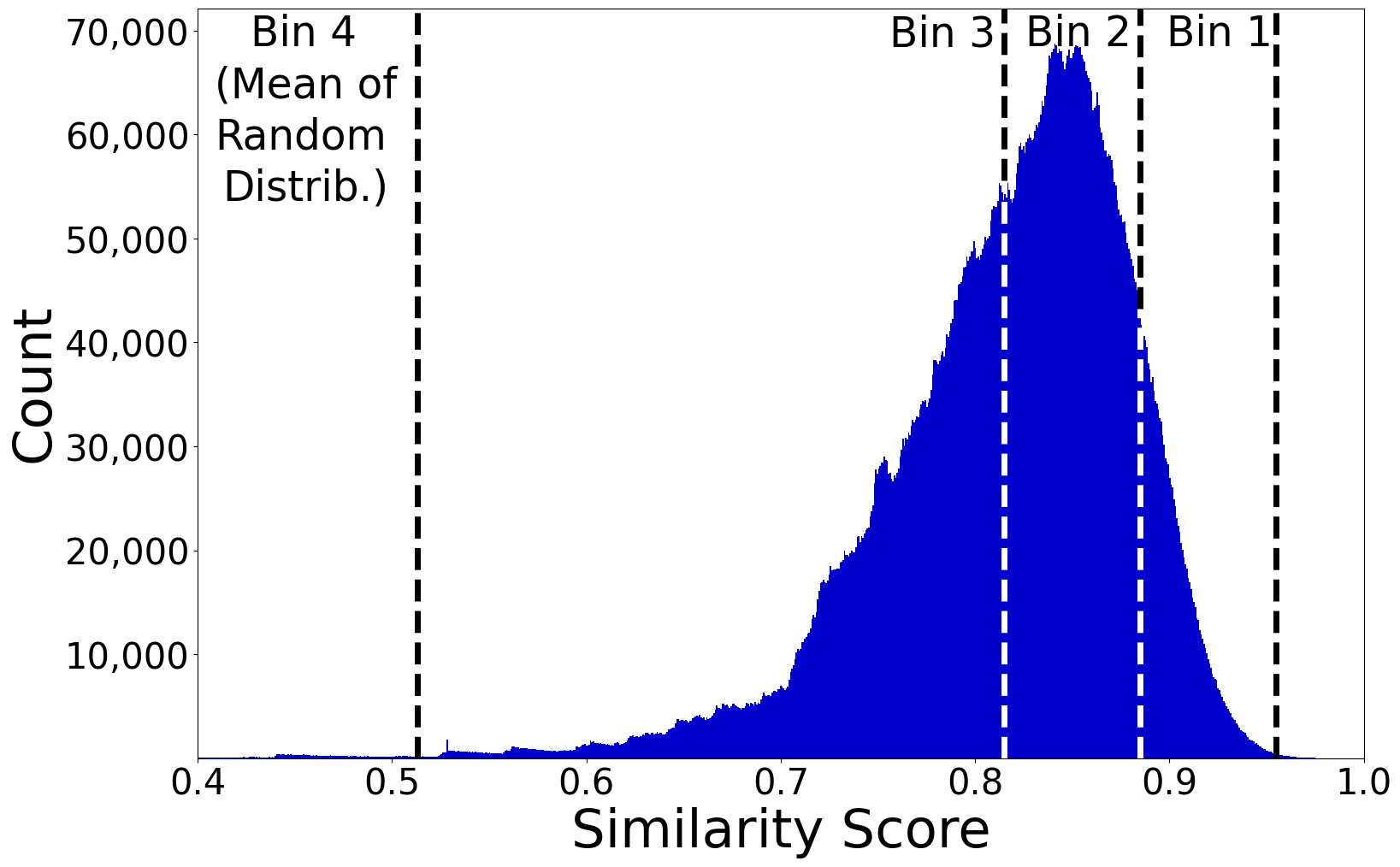}}{} &
\multirow{2}{*}{\backslashbox{\textbf{A}}{\textbf{B}}} &
\textbf{Bin 2} & 
\textbf{Bin 3} & 
\textbf{Bin 4} &
\textbf{Total}\\
& & 
(0.885 $\pm$0.02) & 
(0.815 $\pm$0.02) & 
(0.513 $\pm$0.02) &
(All Trials)\\
\cmidrule(lr){2-6}
  &\textbf{Bin 1} & 
  96.2\% & 
  98.0\% & 
  98.1\% &
  97.4\%\\
 & (0.955 $\pm$0.02) & 
  ($n=156$) &  
  ($n=150$) &  
  ($n=162$) & 
  ($n=468$)\\
 &\textbf{Bin 2} & 
  & 
  73.3\% & 
  93.6\% &
  83.7\%\\
& (0.885 $\pm$0.02) & 
 & 
 ($n=135$) & 
 ($n=141$) & 
 ($n=276$)\\
 &\textbf{Bin 3} & 
  &
  & 
 81.5\% &
 81.5\%\\
 &(0.815 $\pm$0.02) &  
 &  
 &  
 ($n=178$) &
 ($n=178$)\\
   \cmidrule(lr){2-6}
 \multirow{2}{*}{CLMR: Histogram of Top 10k Similar Clips}{} & \multicolumn{5}{c}{\textbf{CLMR Embeddings}} \\
\cmidrule(lr){2-6}
\multirow{9}{*}{\includegraphics[height=3.3cm]{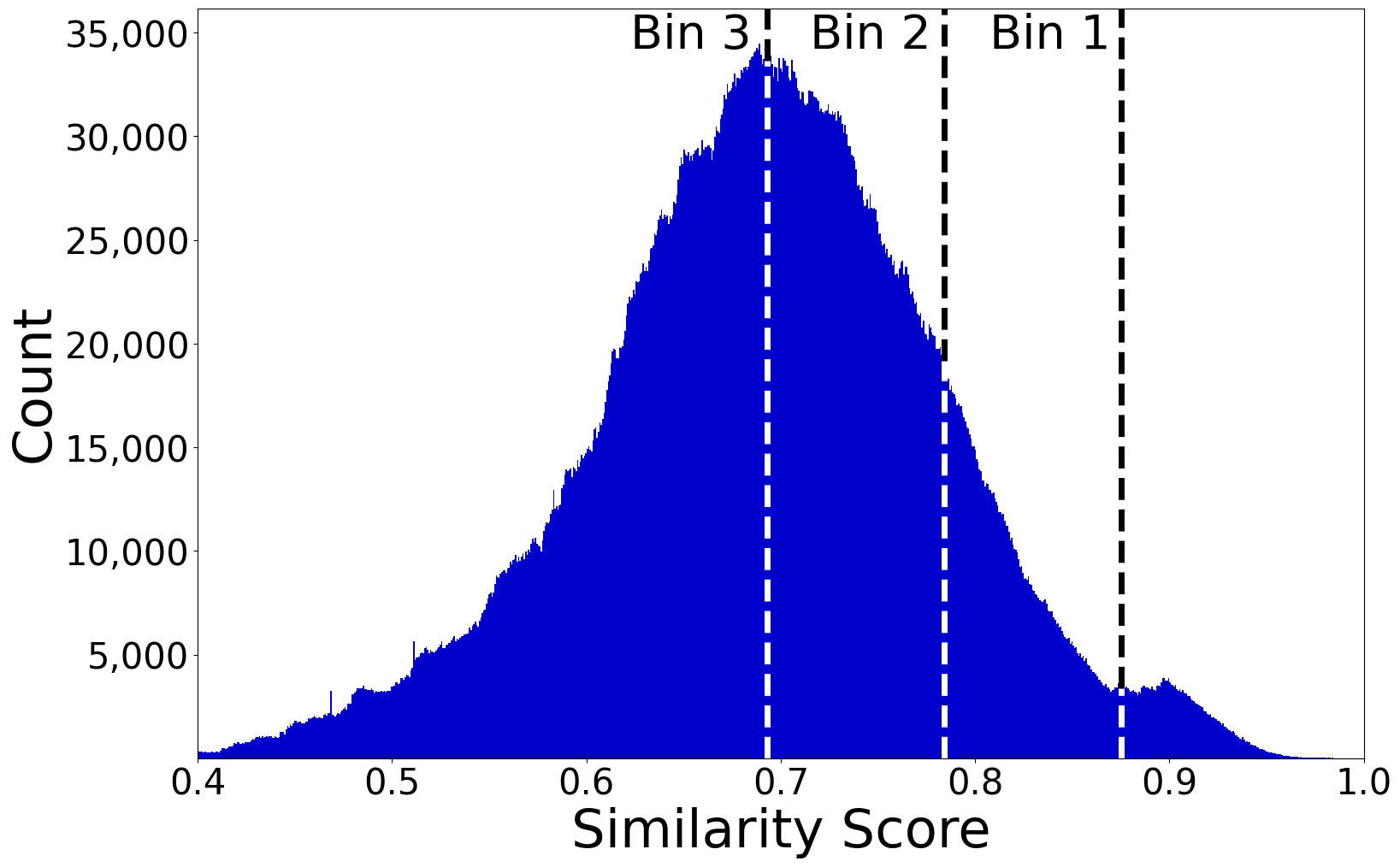}}{} 
&\multirow{2}{*}{\backslashbox{\textbf{A}}{\textbf{B}}} &
\textbf{Bin 2} & 
\textbf{Bin 3} & 
\textbf{Bin 4} &
\textbf{Total}\\
& &
(0.784 $\pm$0.02) & 
(0.693 $\pm$0.02) & 
(0.151 $\pm$0.02) &
(All Trials)\\
\cmidrule(lr){2-6}
 &\textbf{Bin 1} & 
 90.7\% & 
 91.0\% & 
 98.5\% &
 93.2\% \\
 &(0.875 $\pm$0.02) &  
 ($n=150$) & 
 ($n=156$) &  
 ($n=135$) & 
 ($n=441$) \\
& \textbf{Bin 2} & 
 & 
 71.6\% & 
 93.6\% &
 82.7\%\\
& (0.784 $\pm$0.02) & 
 & 
 ($n=155$) & 
 ($n=157$)  & 
 ($n=312$) \\
& \textbf{Bin 3} & 
  &
  & 
 80.7\% &
 80.7\%\\
& (0.693 $\pm$0.02) &  
 &  
 &  
 ($n=140$) &
 ($n=140$) \\
\bottomrule
\end{tabular}
\caption{\label{tab:human_evaluation_results} Table displaying the results form the listening experiment for CLAP \cite{clap_wu_2023} and CLMR \cite{spijkervet2021contrastive} embeddings. Listeners were given a prompt clip ``X'', clip ``A'' with a higher similarity score to it and clip ``B'' with a lower similarity score to it (from within the listed bins). Results show the percent of time listeners rated clip ``A'' (the clips with higher similary scores to the prompt ``X'') as more similar to the prompt clip ``X'' than clip ``B'' (those with lower similarity scores to prompt ``X''). Histograms of the top 10k similar songs, both for CLAP and CLMR, can be found to the left of the table. Bin regions are shown on these histograms. Bin $3$ is centered on the mean of the top 10,000 most similar clips, Bin $2 =+1 SD$, Bin $1 =+2 SD$, and Bin $4$ is the mean similarity score of a randomly selected clip from the entire training data (not just the top 10k).}
\end{table}

Table \ref{tab:human_evaluation_results} contains the results of our listening experiment. We found that human evaluations closely aligned with our quantitative metrics. For both CLAP and CLMR evaluations, listeners affirm, by a wide margin, that clips with higher similarity scores (lower bin numbers) sound more similar to the prompt clip than those with lower scores (higher bin numbers). Clips drawn from the  most-similar bin (Bin 1) to the prompt track ``X'' were rated as more similar to the prompt clip than clips from any other bin 97.4\% of the time for CLAP and  93.2\% of the time for CLMR. For both embeddings, the vast majority of listeners ranked the clips with high cosine similarity to the prompt track (``A'': Bins 1-3) as sounding more similar than the random song (``B'': Bin 4). The comparison with the lowest level of agreement amongst listeners were those comparing Bin 2 (1 standard deviation above the mean of the top 10,000 most similar songs) to Bin 3 (mean of the top 10,000)---this is a much closer comparison mathematically than Bin 3 and 4 (one standard deviation within the top 10,000 versus songs in the top 10,000 compared to a random clip in the top 5 million). For the purposes of this methodology, it is important that the most similar clips (Bin 1) to the prompt sound more similar than any other clip, and that all clips with a high similarity (Bins 1-3) sound more similar than a random clip (Bin 4).

\section{Analysis of Objective Measures}

Our second research question focuses on the effect of different perturbation types and amounts on our methodology's ability to correctly return similar songs. We use these perturbations as a proxy for potential transformations that a generative model may make when outputting a near-duplicate to a song on which it was trained. Then we examine whether the ``vamps'' (music output generated from VampNet) are more similar to the prompt music or to clips from the training data. This sheds light on where the large source of influence comes from: the prompt track or the training data. Then we explore how often generated vamps are highly similar to songs from the training data, leveraging findings from our listening test. 

\subsection{Robustness to Perturbations}\label{sec:perturbation}

We assume that any generative music model will add some degree of variation to a training example during the generation process---the aim of these models are not to replicate the training data exactly. This variation could take many forms such as changing the pitch, speed, melody, etc.. Therefore, in this section we evaluate the ability of our methodology to return target songs that have been modified by given perturbations. For varying amounts of each perturbation, we evaluate how frequently the target song (the unmodified clip) is returned as the most similar, within the top 5 similar songs, and within the top 10 most similar songs. The 7 types of perturbations we evaluate are:
\begin{itemize}
    \item Pitch shift (in semitones; range: -12 to 12)
    \item Time stretch (in \% of song; range: 20\% slower to 20\% faster)
    \item White noise overlaid on top of music (in dB; range: -30 to 30 dB in relation to original audio clip)
    \item ``Mash-up'' of two clips from training data (range: 5/95\% to 95/5\%; e.g., 50/50\%, 60/40\%, etc.)
    \item ``Mash-up'' of one clip from inside and one outside training data (range: 5/95\% to 95/5\%; e.g., 50/50\%, 60/40\%, etc.)
    \item ``Mash-up'' of a prompt clip and the generated vamp (range: 5/95\% to 95/5\%; e.g., 50/50\%, 60/40\%, etc.)
\end{itemize}

We selected these because we envision them as common alterations to music that would not render it unrecognizable by a human listener. We are not seeking to evaluate all types of adversarial noise since we are assuming users and creators are working cooperatively with these generative models to create something novel---not acting maliciously.

\begin{figure}
    \centering
    \includegraphics[width=\textwidth]{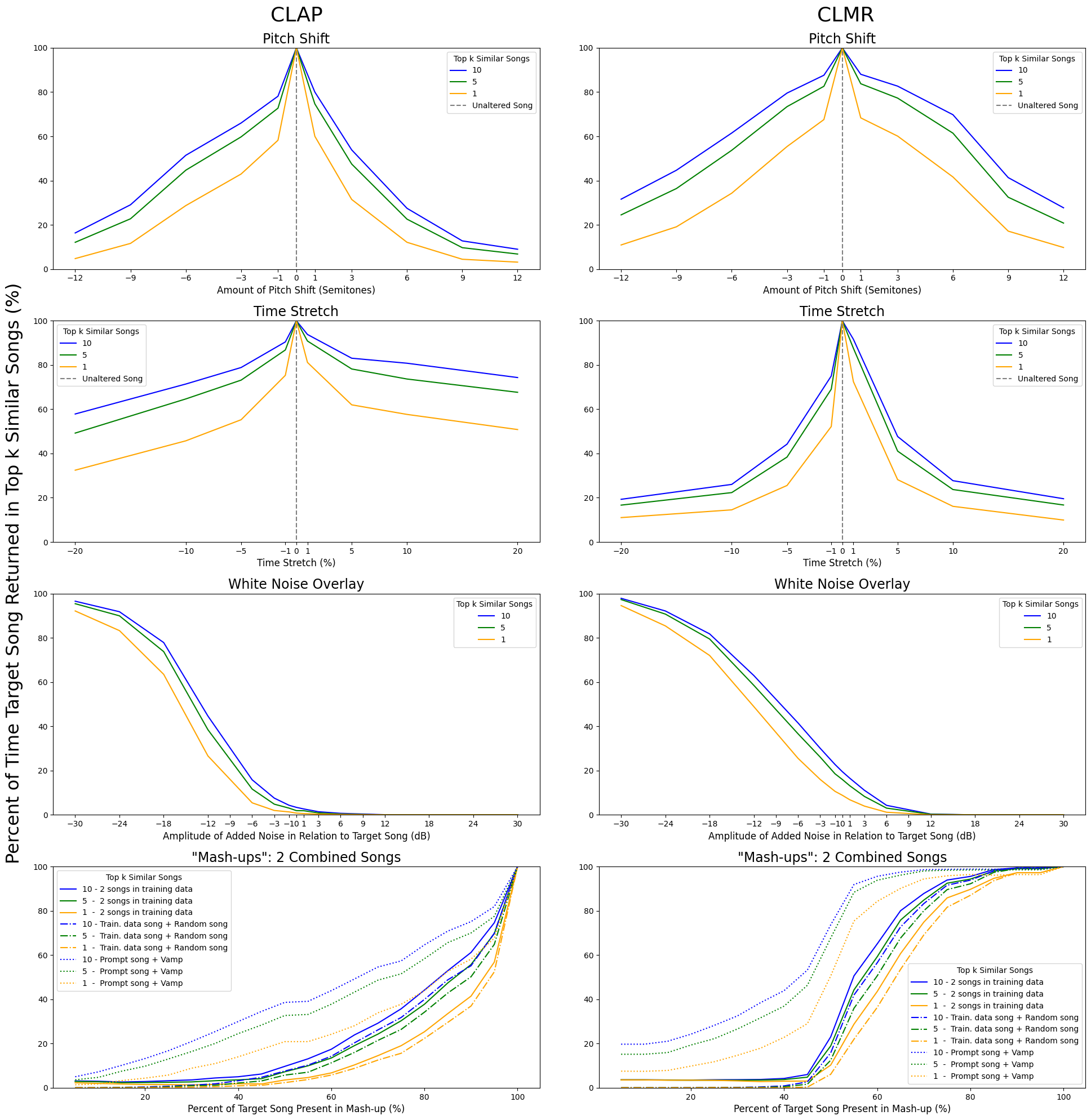}
    \caption{Plots of various amounts of noise perturbations to clips and the percent of the time they were returned in the top $k=10$, $k=5$, and $k=1$ song using our methodology. Analyzed for both CLAP (left column) and CLMR (right column) embeddings. Displays, from top to bottom, pitch shift in semitones, time stretch as percent shortened/elongated, white noise overlay in decibBels to target clip, and mash-ups of 2 songs in training data, 1 song in training data and one random, and a prompt song and its generated vamp.}
    \label{fig:noise-perturbations}
  \Description{Robustness to various noise perturbations.}
\end{figure}

We evaluate all of the audio perturbations for both CLAP and CLMR embeddings to understand the robustness of our methodology while utilizing different embedding networks. All of the results are presented in Figure \ref{fig:noise-perturbations}. For all perturbations except for higher levels of time stretch, using CLMR embeddings is more robust than using CLAP embeddings. Example audio for perturbations is available at \url{https://tinyurl.com/exploring-musical-roots}.

\textit{\textbf{Pitch shift}} is a common perturbation to audio that involves raising or lowering the original pitch of an audio clip without adjusting the length of the clip. For pitch shift, we evaluate shifting the clip between -12 and 12 semitones. A pitch shift of 0 corresponds to an unaltered audio clip. Notably, human perception is extremely robust to pitch shift---no matter in which key ``Happy Birthday'' is sung, everyone knows what song it is. Machines are more susceptible to this perturbation than humans. Both embedding types were robust to small pitch shifts; for changes of $\pm3$ semitones the target song was returned the vast majority of the time. Both embedding types had a lower recall of the target song for larger pitch shifts ($\pm12$ semitones), though CLMR performed much better than CLAP.

\textit{\textbf{Time stretching}} audio clips involves speeding up or slowing down audio while keeping the pitch constant. For this perturbation, we evaluate shifting between making the clip 20\% slower and 20\% faster. A time stretch of 0\% corresponds to an unaltered audio clip. Human perception is extremely robust to both small and large amounts of time stretch in audio (we are able to recognize songs when they are sped up or slowed down). Machines are less adept at recall for large amounts of time stretch. Both embeddings returned the target song for small amounts of time stretch ($\pm1$-$5\%$) a high proportion of the time, but were extremely impacted by larger amounts of time stretch ($>\pm10\%$). Notably, this is the only instance where CLAP embeddings were more robust than CLMR.

\textit{\textbf{White noise overlay}} involves adding randomly generated white noise on top of audio clips. No level of noise added is an unaltered song. We evaluate the level of noise in relation to the amplitude of the original clip in deciBels, ranging from -30 to 30dB (with -30dB being the quietest and 30dB being the loudest we analyzed). To the human ear, $\pm1$ dB is said to be a ``just noticeable difference.'' For context to this analysis, overlaid white noise $\geq18$dB quieter than the song is barely perceptible to the human ear, whereas $\geq18$dB louder noise added makes the music extremely hard to hear. This perturbation has the largest impact on our method's ability to identify the target track. We were only able to consistently return the target song at extremely quiet levels of white noise overlay ($\geq18$dB) that were barely perceptible to the human ear. Luckily, this is not an anticipated common level of noise added by a generative model: generative music models will attempt to add more ``musical'' noise and variation to songs rather than random white noise. 

\textit{\textbf{``Mash-ups'' of two combined songs}} are defined here as splicing two clips together at different percentage levels (e.g., for 50/50\% the first 1.5 seconds are the target song A and the last 1.5 seconds are some other song B; for 75/25\% the first 2.25 seconds are target song A and the last 0.75 seconds are some other song B). We evaluate three different types of ``mash-ups'': (1) combining two clips from the training data, (2) combining one clip from the training data with a song from outside of the training data, and (3) combining a prompt track and its generated vamp from VampNet.

For each mash-up, we seek to identify the percent of time the target track ``A'' is returned in the top similar songs when it is present in a given proportion of the mash-up (e.g., 5\%-95\% as indicated by the x-axis). For (1) combining two songs from the training data, we only seek to identify one target song ``A'', for (2) one song from the training data and one random song, we only seek to identify the target song from the training data, and for (3) one prompt song and its corresponding vamp, we only seek to identify the prompt song.

This is the most interesting disparity between the robustness of CLMR and CLAP embeddings. For each of the three types of mash-ups, CLMR embeddings only need 50-60\% of the target song present in the mash-up to consistently return it in the top similar songs, but CLAP embeddings need more than 80\%. For both embeddings, at each mash-up proportion the model was able to return the target song (prompt song) for mash-ups with vamps more consistently than for combining two different songs. This indicates that the vamp is more similar to the prompt song than two randomly selected songs are to each other. The goal of this perturbation is to see if combining these target songs with their prompt affects the method's ability to identify the target song as the most similar. As can be seen in Figure \ref{fig:noise-perturbations}, when the majority of the song analyzed is the vamp (e.g., x-axis $\le50\%$), it does not return the target (prompt) but rather other songs in the training data; this is explored further in the following section.

\subsection{Systematic Evaluation of Generative Music Model}\label{sec:overall_eval_big_model}

\begin{table}[ht]
\begin{tabular}{@{}cc|c|c|c|c@{}}\toprule
  \multicolumn{6}{c}{\textbf{Descriptive Statistics: Systematic Evaluation of Generated Music (Vamps)}} \\
\midrule
 &&
 \multicolumn{2}{c|}{\textbf{Full Model}} &
  \multicolumn{2}{c}{\textbf{Small Training Dataset Model}}\\
\midrule
 \multicolumn{2}{c|}{\multirow{2}{*}{\textbf{Similarity Score}}{}} & 
 \multirow{2}{*}{Vamp \& Prompt}{} &
Vamp \& &
 \multirow{2}{*}{Vamp \& Prompt}{} &
Vamp \&  
\\
&
& 
&
\#1 Similar Track &
 &
 \#1 Similar Track\\
\midrule
\multirow{4}{*}{\textbf{CLAP}}{} & 
Mean     & 0.393 & 0.795 & 0.413 & 0.804\\
& Median   & 0.402 & 0.815 & 0.425 & 0.827\\
& St. Dev. & 0.151 & 0.084 & 0.150 & 0.081\\
& Max      & 0.855 & 0.960 & 0.836 & 0.950\\
\midrule
\multirow{4}{*}{\textbf{CLMR}}{} & 
 Mean     & 0.166 & 0.846 & 0.183 & 0.850\\
& Median   & 0.153 & 0.850 & 0.175 & 0.853\\
& St. Dev. & 0.189 & 0.054 & 0.190 & 0.051\\
& Max      & 0.853 & 0.999 & 0.853 & 0.999\\
\bottomrule
\end{tabular}
\caption{\label{tab:descriptive_stats_full_eval} Descriptive statistics of systematic evaluation of VampNet generative music model. Generated pieces of music (vamps) are consistently less similar to the prompt song provided to the model at generation time than they are to other music from the training data. Between the model trained on a smaller training set and the one trained on the larger set, there is not a large difference for the closest clip in the training data or similarity to the prompt, though the model trained on the smaller training dataset shows slightly more similarity in both cases.}
\end{table}

A primary goal of this work is establishing a methodology that enables users and creators of generative music models to understand the attribution of the training data for a generated piece of music. As a case study, we systematically evaluate VampNet \cite{garcia2023vampnet} to demonstrate how to employ this technique to understand training data attribution on both individual songs and an entire model. To do this, we generate 10,000 ``vamps'' (music clips generated by VampNet) from 1,000 10-second prompt clips (10 different vamps per clip), and evaluate the most similar clips in the training data to the vamps. We embed each of the 10,000 vamps as a feature vector using both CLMR and CLAP embeddings for this analysis and for both embedding networks analyze the most similar 50 clips by cosine similarity (out of the five million+ in our vector store). Examples of highly-similar clips are available at  \url{https://tinyurl.com/exploring-musical-roots}.

For each of the 10,000 vamps, the prompt that generated the vamp was rarely among the top similar clips returned by our methodology. This is by VampNet's design---this generative model seeks to make variations and new pieces of audio, not direct copies of the prompt. However, this music did not come out of thin air and we seek to understand the attribution of the \textit{rest} of the training data on the new generations. For CLAP embeddings, the average cosine similarity between a prompt clip and the generated vamp was 0.393, ($sd=0.151$), whereas on average, the closest clip had a similarity score of 0.795 ($sd=0.084$). CLMR had a similar disparity; the full descriptive results can be found in Table \ref{tab:descriptive_stats_full_eval}.

In order to understand the direct effect of training data size on training data attribution, we repeated the above analysis on a version of VampNet trained on a much smaller dataset. The original form of VampNet released by Flores Garc\'ia et al. in 2023 (and used in the above experiment in Section \ref{sec:overall_eval_big_model}) had 333M and 275M parameters for the coarse and coarse-to-fine models, respectively, and was trained on 795,000 songs. For this analysis, we retrain VampNet with the same architecture and number of training steps, simply with a different sized training data. We only train this version of the model using 30,550 songs. With that one change, we recreate the analysis in the above section (\ref{sec:overall_eval_big_model}) and identify the impact of a smaller training dataset on the overall training data attribution. We provide the findings in the last two columns of Table \ref{tab:descriptive_stats_full_eval}, and as you can see they are quite similar to that of the model trained on the full training data. We perform every analysis in this section on both the full model and the model trained on the smaller dataset, but found no meaningful differences, so for the remainder of this section we will only present results for the larger model.

As noted in the above analysis on this methodology's robustness to various perturbation in Section \ref{sec:perturbation}, our methodology utilizing CLMR embeddings is more robust to perturbations of ``mash-ups'', or combining elements of new clips with clips present in the training data and still returning the original clip as the one with the highest similarity score. Thus for the remainder of this section we will focus on CLMR embeddings for this case study using VampNet.

Leveraging insight from our listening study (Section \ref{sec:experiment}), human evaluation affirms that within CLMR embeddings, music clips with a similarity score of 0.875 or higher sound significantly more similar than clips with lower similarity scores. For this analysis, we utilize that same top bin as a benchmark and evaluate how often the most similar clips have similarity scores at or above 0.875. Findings are presented in Table \ref{tab:clmr_high_similarity_scores}. Over 30\% of the vamps generated had at least one song with a similarity score $\geq0.875$. Looking at scores in 0.02 increments above this benchmark similarity score, almost 20\% of vamps had at least one song in the training data with a similarity score $\geq0.895$, 9\%  $\geq0.915$, 3\% $\geq0.935$, and almost 1\%  $\geq0.955$. Looking more broadly at the top 5 songs and top 10 songs, it is evident that the songs with these high similarity scores were concentrated among the most similar 1-2 clips, as opposed to having the entirety of the top 10 most similar clips have extremely high similarity scores. This indicates that at least 30\% of the time, small sets of  songs from the training data were highly influential on generated vamps rather than a combination of many songs.

\begin{table}[ht]
\begin{tabular}{@{}c|c|c|c|c|c|c@{}}\toprule
  \multicolumn{7}{c}{\textbf{Vamps with Highly Similar Songs in Training Data}} \\
\midrule
 &
 \multicolumn{2}{c|}{\textbf{\#1 Most Similar Songs}} &
 \multicolumn{2}{c|}{\textbf{Top \#5 Most Similar Songs}} &
 \multicolumn{2}{|c}{\textbf{Top \#10 Most Similar Songs}} \\
\midrule
 \multirow{2}{*}{\textbf{Similarity Score}}{} & 
 $k$-clips &
 \multirow{2}{*}{\% of Total}{} &
 $k$-clips &
 \multirow{2}{*}{\% of Total}{} &
 $k$-clips&
 \multirow{2}{*}{\% of Total}{} \\
 &
 (n=10,000)&
 &
 (n=50,000)& 
 &
 (n=100,000)\\
\midrule
$\geq0.955$ & 89    & 0.89\%  & 199     & 0.40\%  & 254    & 0.25\% \\
$\geq0.935$ & 317   & 3.17\%  & 846     & 1.69\%  & 1,223  & 1.22\% \\
$\geq0.915$ & 924   & 9.24\%  & 2,458   & 4.92\%  & 3,139  & 3.14\% \\
$\geq0.895$ & 1,929 & 19.29\% & 5,956  & 11.91\%  & 8,786  & 8.79\% \\
$\geq0.875$ & 3,201 & 32.01\% & 10,975 & 21.95\%  & 17,291 & 17.29\%\\
\bottomrule
\end{tabular}
\caption{\label{tab:clmr_high_similarity_scores} For 10,000 vamps generated from VampNet, a table displaying the number and percent of the time the top $k=1$, $k=5$, and $k=10$ most similar songs were at or above given similarity scores for CLMR embeddings. The lowest similarity score in this table (0.875) corresponds to the score from the subjective evaluation (Section \ref{sec:experiment}) that human evaluators asserted sounded more similar than any score below.}
   \vspace{-1.5mm}
\end{table}

\section{Discussion}\label{sec:discussion}

These findings establish that the framework we propose is an effective means to systematically evaluate the training data attribution on any generative music model. This method is replicable and should be employed by creators of generative models prior to deployment so they are able to have a greater understanding of their outputs. If exposed to end users, this framework also enables anyone to verify if they are copying music and learn about influences of their ``novel'' generations. Based on our analysis and experimentation, we next elaborate on limitations of our specific approach, identify new avenues of research that could be pursued, align with an ethical taxonomy of harms of generative audio models \cite{barnett2023ethical}, and elaborate on how this framework can be utilized to help both model creators and end users.

The authors first acknowledge the limitations of this approach. First, the scope is intentionally limited to exclude lyrics (see Section \ref{sec:scope}). We do this because lyrics fall more within the text domain and we wanted to focus on the overall sound of the generation. As generative music models continue to progress this can become an extremely important area of memorization and copyright infringement, and we encourage future research to examine lyric memorization in tandem with our approach. Additionally, our scope did not include any individualized feature levers for similarity such as singling out when two pieces of music having the same timbre, melody, or rhythm. We did this to both focus on a low-burden implementation for future model creators who would follow this methodology as well as to identify encompassing interacting similarities without isolating any musical feature. However, these could be extremely useful for both model creators and users seeking to understand their generations at a fundamental level, and having the ability to learn which features of songs are replicated and similar is an important area of future research. This has also been suggested by Lee et al. \cite{lee2020disentangled} as interaction paradigms that allow users of these models to identify dimensions of music that are of interest for different music retrieval tasks.

The authors chose to utilize an embedding-centered approach in order to maximize the ease for model creators to implement this methodology. There are also other methods for examining memorization such as mel spectrograms \cite{bralios2023generation}, and these could identify similarities that embeddings may not uncover, potentially at the expense of ease of implementation---future work should examine how to integrate these methods. Also inherent to this embedding-centered approach is the choice of which embeddings to implement. We experimented with a variety of different embeddings, ultimately selecting CLAP and CLMR to validate and align with human subjective evaluation (section \ref{sec:experiment}). However, there are and will continue to be more embeddings out there that can offer various advantages---we leave it up to the model creator to select their own preferred embeddings.

During our analysis of VampNet we also identified highly influential songs by examining clips that were highly similar to an abnormally large percent of generated vamps. Some songs were present in more than 2\% of the top 10 songs for generated vamps, indicating that the model had a preference for certain musical representations over others. An interesting avenue for further research is to focus on these highly influential songs and analyze why these songs are so favored by the model. Some creative work could center on fine tuning a model that either focuses on more novelty or conversely design a model that intentionally reproduces ``the hits'' in its generations. 

A recent evaluation of the ethical implications of generative audio models \cite{barnett2023ethical} identified several potential broader negative impacts of generative music models including  potential for cultural appropriation, copyright infringement, and loss of agency and authorship of the creators. Our work aims to combat these issues both at the time of output generation and prior to model release. Our method enables creators and users to understand training data influences on model output. This can prevent cultural appropriation by giving users the opportunity to engage with the influences of the music, and potentially prevent copyright infringement if the user realizes the generated piece of music is too similar to the identified influences.

A potential use of this training data attribution is identifying the contributions of creators of the training data (i.e., musicians), and compensating them for generations. If a generative model starts generating income (e.g., through subscriptions such as OpenAI) or users monetize their outputs, they can use this tool to understand varying attribution of the underlying training data and compensate the original creators accordingly.

Our proposed framework is easy to implement for any generative model for which there is access to the full training data. As the training data scales, the only corresponding architecture scaling is (1) compute-time for embedding the training data and (2) the vector database storage required to hold the dataset. Once uploaded to a vector store, performance and retrieval time scale efficiently. Every model creator should consider using a framework similar to ours as a means to understand proper attribution of training data.

\section{Conclusion}

In this paper we proposed an easily implementable framework for creators of generative music models to  evaluate training data attribution. It can be used to prevent appropriation, copyright infringement, and otherwise uninformed creations, enabling model creators and users to understand the influences of their generated outputs by systematically identifying similar songs in the training data. We evaluated an objective measure of cosine similarity for two embeddings and verified that they align with human perception with a subjective listening test. We also evaluated how robust our framework is to various forms of perturbations we anticipate models adding to training data during the transformation to ``novel'' output. We perform a case study on VampNet \cite{garcia2023vampnet} in order to validate the efficacy of our framework as well as elucidate some potential findings that utilization of our method can provide. This work is a step towards transforming a generative model from a crutch replacing artistic knowledge to a tool creators and users alike can use to become better and more informed artists.

\section{Research Ethics and Social Impact}

The authors of this paper took the ethical considerations and social impact of this work seriously. A recent exhaustive study of the ethical implications of generative audio models \cite{barnett2023ethical} found that less than 10\% of research on generative audio models published discussed any sort of potential negative impact of their work. We took that as inspiration to center our work around the ethical concerns and attempt to build a bridge between ethicists and generative audio engineers. 

As mentioned in the discussion (Section \ref{sec:discussion}), among the negative impacts uncovered for generative music models were the potential for cultural appropriation, copyright infringement, and loss of agency and authorship of the creators. This work aims to combat these issues at the time of generation, on a track by track level. By uncovering the roots of a given piece of generated music, we can empower the user of the model to understand where the music came from and learn about the influences. 

A primary concern the authors have for this work is that future model creators will simply use this framework as a checkbox to complete their ethical evaluations. They may use this framework and assume since they did so, there are no other potential societal impacts or ethical harms to consider in regard to generative music models. This work only tackles a portion of the issues, and is only a first step in doing so. Though our method can highlight instances of copyright infringement and cultural appropriation, it by no means will catch everything. Though this can assist with educating users about the influences of their work, it will not solve the potential loss of agency and authorship users and musicians could feel when using these models. It does nothing to address creativity stifling, predominance of western bias, overuse of publicly available data, non-consensual use of training data, or job displacement and unemployment. It also requires energy consumption to generate the embeddings and perform searches, so it contributes to the issue of energy consumption of generative models rather than combating it.

The authors of this work are both musicians and generative AI researchers, so we approached this work from the perspective of attempting to improve generative music models as tools of co-creation with musicians. Our backgrounds as generative AI enthusiasts shaped our mentality that generative music models are a positive contribution to the art scene and society, so we may be blind to potential shortcomings of these models beyond the ones listed above. We also acknowledge that we are a small subset of human beings, so our opinions are by no means representative of either musicians or generative AI researchers, though we believe we are uniquely positioned at the intersection of music and AI. Additionally, we are all from the western hemisphere so our work contributes to the predominance of western bias in generative music. 

In regard to the experiment utilizing human evaluators to subjectively analyze similar pieces of music, we ensured that our study was in line with institutional review board standards (our study was determined to be exempt). We had a thorough consent form for the crowdworkers and ensured they knew they could quit at anytime without any sort of penalty. We timed ourselves taking the survey and attempted to pay them a fair wage (estimated \$22.50 per hour, higher than any minimum wage in the United States). We even paid users  who failed the listening pre-screening test for their time and thus were not able to take our survey, even though they did not contribute data to our study. However, we acknowledge that ethical crowdsourcing goes beyond fair pay \cite{shmueli2021beyond, schlagwein2019ethical}, and tested the listening test thoroughly prior to launch to be certain there would be no burden to crowdworkers beyond potential boredom. The most sensitive data we had access to were the Mechanical Turk IDs of users, but we held these on secure servers.

The authors determined that the positive impact of this work outweighed these potential harms, especially since the primary motivation of this work is to address a few existing ethical issues in generative audio. However, it is essential to acknowledge these potential risks and where our method falls short.
 
\begin{acks}
The authors would like to thank Max Morrison for his help with deploying Reproducible Subjective Evaluation (ReSEval) \cite{morrison2022reproducible}, as well as his helpful comments.    
\end{acks}

\bibliographystyle{ACM-Reference-Format}
\bibliography{references}

\begin{appendix}

\section{Appendix}\label{sec:appendix}

\subsection{Search and Vector Database}\label{sec:appendix-vector-database}

In order to quickly and efficiently query against a sufficiently large set of training data, we use a vector database to hold those embeddings. For this we utilized Pinecone, a vector database that is highly scalable for large stores of embeddings. Pinecone is built on top of Facebook AI Similarity Search (Faiss) \cite{johnson2019billion}, an open source library built to implement extremely efficient search for \textit{k}-selection. In order to quickly return queries (we currently have 5,000,000 and it takes milliseconds to return the top \textit{k} vectors), the vector database employs MinHashing locality sensitive hashing. MinHashing compresses sparse vectors into densely packed number signatures, and a banding approach to locality sensitive hashing \cite{indyk1998approximate} quickly identifies candidate pairs to perform similarity search on rather than an exhaustive search through the whole dataset. In addition to these methods, Pinecone also uses Hierarchical Navigable Small World graphs for indexing the vectors to increase search speed, and then product quantization in order to compress the vectors in order to use less storage and dramatically speed up the search without sacrificing accuracy. There are many other vector databases available, such as Faiss \cite{johnson2019billion}, Milvus \cite{wang2021milvus}, and Weaviate; any of these will work for our framework.

\subsection{Subjective Evaluation Example Question}\label{sec:reseval-example}

In order to evaluate our objective metric with subjective human perception, we performed a listening evaluation on both CLAP and CLMR embeddings. After passing a pre-screening listening test, each evaluator was asked to perform 10 pair-wise evaluations of two 3-second audio clips and determine which they thought sounded more similar to the prompt. The selection of 10 pair-wise evaluations was randomly distributed among all participants. The order of clip ``A'' (the song with a higher similarity score) and clip ``B'' (the song with a lower similarity score) was randomized for every evaluation. Below is an example of what an evaluator would see.

\begin{figure}[ht]
    \centering
    \includegraphics[width=\textwidth]{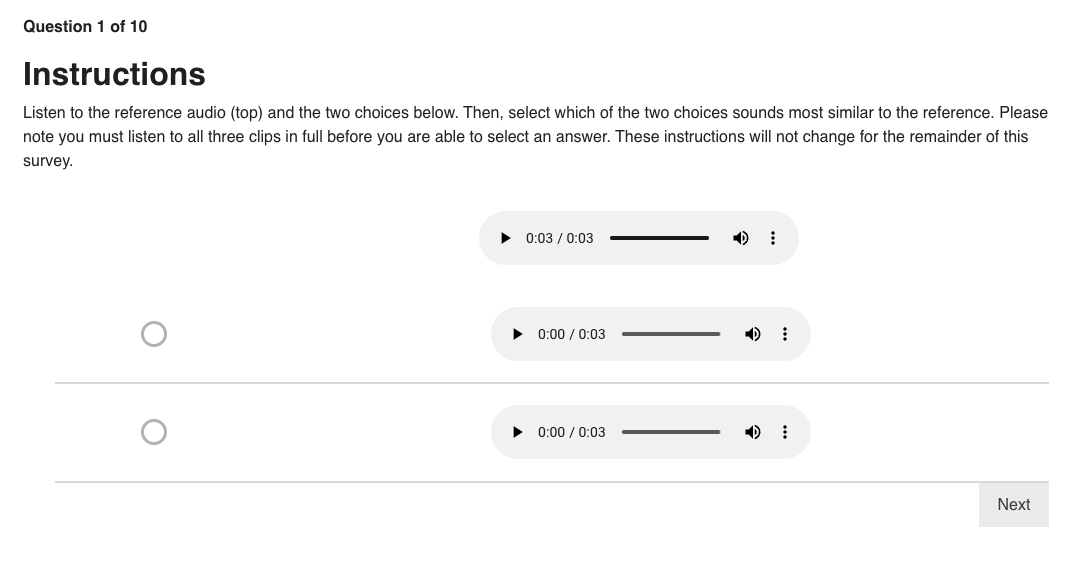}
    \caption{Example question participants had in our subjective evaluation. }
    \label{fig:reseval_example_pic}
  \Description{Example of a question on the subjective listening evaluation.}
\end{figure}

\end{appendix}

\end{document}